\documentclass[12pt,showpacs,preprintnumbers,superscriptaddress,nofootinbib]{revtex4}
\usepackage{amsfonts}
\usepackage{amssymb}
\usepackage{amsmath, graphicx}
\usepackage{dcolumn}
\usepackage{bm}
\usepackage{epstopdf}
\usepackage{color}
\usepackage{soul}
\usepackage{cancel}

\newcommand{\hs}{\hspace*{0.3cm}}

\newcommand{\be}{\begin{equation}}
\newcommand{\ee}{\end{equation}}
\newcommand{\bea}{\begin{eqnarray}}
\newcommand{\eea}{\end{eqnarray}}
\newcommand{\ben}{\begin{enumerate}}
\newcommand{\een}{\end{enumerate}}
\newcommand{\bit}{\begin{itemize}}
\newcommand{\eit}{\end{itemize}}
\newcommand{\bde}{\begin{widetext}}
\newcommand{\ede}{\end{widetext}}

\newcommand{\crn}{\nonumber \\}

\newcommand{\al}{\alpha}
\newcommand{\la}{\lambda}

\newcommand{\ga}{\gamma}

\newcommand{\pa}{\partial}

\newcommand{\fr}{\frac}

\newcommand{\bc}{\begin{center}}
\newcommand{\ec}{\end{center}}
\newcommand{\Ga}{\Gamma}
\newcommand{\de}{\delta}

\newcommand{\ep}{\epsilon}

\newcommand{\si}{\sigma}

\input{tcilatex}

\usepackage{soul}
\usepackage{cancel}

\input{tcilatex}

\begin{document}

\title{\boldmath 
Prospects for detecting $SU(2)_L$ hidden sector bosons  at ILC}

\author{D. T. Binh}\email{dinhthanhbinh3@duytan.edu.vn}
\affiliation{Institute of Theoretical and Applied Research, Duy Tan University, Hanoi 100000, Vietnam}

\author{H. N. Long}
\email{hnlong@iop.vast.ac.vn}
\affiliation{Institute of Physics,   Vietnam Academy of Science and Technology, 10 Dao Tan, Ba
 Dinh, Hanoi, Vietnam }


\date{\today }

\begin{abstract}

 We investigate the possibility of detecting hidden vector gauge bosons at ILC linear collider. The study is performed in the framework of hidden sector extension of Standard Model with 3 degenerate dark gauge bosons. By studying the cross section of pair dark gauge boson with photon at initial state radiation we found that at the energy $\sqrt{s} \approx$ 1300 GeV the cross section can be as large as $48fb$, the same order  ($\mathcal{O}(fb)$) with the  irreducible background  of the Standard Model. Hence  more methods needed to be done to eliminate the background for this model.

	\end{abstract}
	
\pacs{12.60.-i,95.35.+d
}
\maketitle

\textbf{Keywords}: Extensions of
electroweak Higgs sector, Dark Matter

\section{Introduction}
\label{intro}

Despite no evidence of new physics signal has been observed at the colliders yet,
the observations from cosmology reveal that more than 20$\%$ of the whole Universe is made up by the so-called dark matter (DM) \cite{WMAP,Obv-Sig-DM1,Obv-Sig-DM2,Obv-Sig-DM3,Obv-Sig-DM4}. It is clear that DM problem can not be addressed in the framework of Standard Model (SM). This situation motivated a variety of efforts in theoretical extension models of SM such as supersymmetry\cite{Theory-DM1}, extra-dimension \cite{Theory-DM2,Theory-DM3} little Higgs \cite{Theory-DM4} and 3-3-1 models \cite{331-1,331-2,331-3,331-4,331-5,331-6,331-7,331-8,331-9,331-10,331-11,331-12,331-14}. On the experiment side, there are  searches for DM through non-gravitational interactions such as direct, indirect \cite{Status-DM-searches} and collider signatures \cite{LHC-DM1,LHC-DM2,LHC-DM3, Collider-overview}.

There are many possibility for candidate of  DM such as weakly interacting massive particles (WIMPs) \cite{WIMP-1}, super WIMP\cite{sWIMP1,sWIMP2,sWIMP3,sWIMP4,sWIMP5}, asymmetric DM \cite{Asym-DM1,Asym-DM2,Asym-DM3,Asym-DM4,Asym-DM5} or  QCD axion \cite{ADMX,QCD-Axion-Graham} etc. However, DM may have origin   from hidden sector\cite{hidden-Sector1,DarkSM-1,DarkSM-2,DarkSM-3} etc. In this approach DMs are not particles with SM quantum number and there exists a portal connecting the visible
sector to the hidden sector.

WIMPs  with mass of order  $\mathcal{O}$(100) GeV  are among the favoured candidate particles to provide the observed cosmological abundance of DM. Collider experiment can complement to direct and indirect detection experiments which look for signals of WIMPs. Since the WIMPs leave collider experiments
undetected due to their weak interaction with matter. However, the undetected signal or the missing transverse energy ($E\!\!\!/_T$) can also come from hidden sector DM. Collider searches typically rely on
$E\!\!\!/_T$ appearing in cascade decays of more heavy exotic particles or through a portal, thus assuming a specific extension of the SM of particle physics.

The process $e^- e^+ \rightarrow \ga  + "E\!\!\!/_T"$ is an efficient way to explore new
physics. One of the great advantages of the $e^+ e^- $ collider  e.g.  ILC is its clean environment  whose
initial state is exactly known. This knowledge, in conjunction with the  momentum and energy resolution and the high luminosity of the ILC detector, reduces the systematic uncertainties on the measurements. The ILC experiment therefore provides all the requisites for high precision measurements within the SM and physics beyond.

 In the SM, the missing energy can be carried by neutrino which comes
from the $Z$-exchange in the $s$-channel and from the $t$-channel through $W$ exchange.
Beyond SM, the $E\!\!\!/$ could come from new generation of neutrinos,
radiative production of new WIMP or new particles.

In this work, we will investigate the dark sector extension of the SM \cite{DarkSM-1}. Besides a small number of new parameters,
the model has an important custodial symmetry which leads to the degeneracy of the mass of the 3 dark gauge bosons. Since all other particles are $SO(3)$ singlets, any decay of the gauge bosons is forbidden by the custodial symmetry \cite{DM-Stability}. Thus these dark gauge bosons are stable as required for candidate for DM.

	The paper is organized as follows: In section \ref{sec2} we will briefly review the dark sector extension of the SM. In section III we evaluated the cross section for the process $e^-e^+ \rightarrow \ga$ + "Dark matter" in the model under consideration. Section IV is dedicated for numerical analysis and finally conclusions in section V.

\section{ Review $SU(2)_L$ with dark sector  model}
\label{sec2}

The non-Abelian gauge dark matter model was proposed in several papers \cite{DarkSM-1,DarkSM-2,DarkSM-3}. In this paper we will focus on the model \cite{DarkSM-1}. This model is based on two assumptions. First, the existence of a $A^\mu$ gauge multiplet associated to new non-Abelian gauge symmetry $G^D$.  All SM particles are singlets under $G^D$. Second,  the non-Abelian field only interacts with  the SM  through a complex scalar Higgs portal field, $\phi$, which is singlet of the SM but charged under $G^D$. Mixing of $A^\mu$ with the SM gauge bosons  is forbidden by the non-Abelian character of the extra gauge symmetry. Assuming $G^D= SU(2)$, then the most general Lagrangian  is given as:
\bea
{\cal L}&=& {\cal L}^{SM} -\fr{1}{4} \left(F'_{\mu\nu}\right)^2
+(D_\mu \phi)^2- \la_m \phi^\dagger \phi H^\dagger H-\mu^2_\phi \phi^\dagger \phi -\la_\phi (\phi^\dagger \phi)^2
\label{VDML}
\eea
with $D_\mu \phi=\pa_\mu \phi - i\fr{g_\phi}{2} \si  A'_\mu$ where  in the SM Lagrangian  the Higgs potential contains:
$ V^{SM}_H = \mu^2 H^\dagger H + \la (H^\dagger H)^2$ with $H^T=(H^+, H^0)$.


If  the gauge group $G^D$ of the hidden sector is spontaneously broken by VEV  $\langle \phi \rangle = v_\phi$, and making the following rotations
\bea
\phi^T &= &e^{(i \si\cdot\xi/v_\phi)} \cdot (0, \,\fr{1}{\sqrt{2}} [v_\phi+\eta'] )\, ,\label{r1}\\
H^T &= & e^{(i \tau \cdot \zeta/v)}\, (0, \,\fr{1}{\sqrt{2}} [v+h'] ) \, ,
\label{r2}
\eea
where $v=v_{SM}=246$ GeV. Within rotation (\ref{r1}) the gauge boson $A'_\mu$ transform into $A_\mu$ while transform in Eq.(\ref{r2}) leads to the ordinary
unitary gauge where the diagrams containing Goldstone boson are ignored. If so the gauge bosons in the hidden sector get mass given by
\be
m^2_A = \fr 1 4 (g_\phi v_\phi)^2 \, .
\label{r3}
\ee
Then ones get final expression \cite{DarkSM-1}

\bea
{\cal L}&=&{\cal L}_{SM}-\fr 1 4  (F_{\mu \nu})^2  + \fr 1 8  g_\phi A_\mu \cdot A^\mu \eta'^2
+\fr 1 4  g_\phi v_\phi A_\mu \cdot A^\mu \eta' +\fr 1 2 (\partial_\mu \eta')^2 \crn
&-&\fr{\la_m}{2}(\eta'+v_\phi)^2 H^\dagger H
- \fr{\mu_\phi^2}{2} (\eta'+v_\phi)^2   - \fr{\la_\phi}{4} (\eta'+v_\phi)^4 \, ,
\label{r4}
\eea
where the terms in the second line of Eq.(\ref{r4}) are considered as new scalar potential.

The total scalar potential being  sum over the mentioned terms and $V_{SM}$ has the constraint conditions as follows \cite{DarkSM-1}
\bea
v^2 &= &\fr{-\mu^2 \la_\phi +\fr{1}{2}\la_m \mu^2_\phi}{\la \la_\phi-\fr{1}{4} \la_m^2} \, ,\crn
v_\phi^2 & = & \fr{-\mu_\phi^2 \la +\fr{1}{2}\la_m \mu^2}{\la \la_\phi-\fr{1}{4} \la_m^2}\,.
\label{r5}
\eea
In the $(h', \,\eta')$ basis, the square mass matrix is given as
\be
M^2=\left(
\begin{array}{ cc}
   m^2_{h'}&  m^2_{h' \eta'}\\
    m^2_{h' \eta'} & m^2_{\eta'}
\end{array}
\right) \,,\,
\label{massmatrix}
\ee
with
\be
m_{h'}^2=\fr{-2 \mu^2 \la \la_\phi + \la \la_m \mu^2_\phi }{\la \la_\phi -\fr{1}{4}\la_m^2} \, \hspace{1cm}
m_{\eta'}^2= \fr{-2 \mu_\phi^2 \la \la_\phi + \la_\phi \la_m \mu^2 }{\la \la_\phi -\fr{1}{4}\la_m^2} \, \hspace{1cm}
m^2_{h'\eta'}=\la_m v v_\phi \,.
\ee
The physical states are

\be
\left(
\begin{array}{ c}
   h'\\
   \eta'
\end{array}
\right) = \left(
\begin{array}{ cc}
  \cos \beta&   \sin \beta \\
   -\sin \beta & \cos \beta
\end{array}
\right) \,\,
\left(
\begin{array}{ c}
   h\\
   \eta
\end{array}
\right)\,,
\ee
where $\tan 2 \beta= 2 m^2_{h'\eta'}/(m^2_{\eta'}-m^2_{h'})$.

Within the above transformation, the SM part is not affected and final
 Lagrangian in the physical state basis is given by \cite{DarkSM-1}


\bea
\mathcal{L}&=& -\fr 1 4 (F_{\mu \nu})^2
+\fr{1}{8} (g_\phi v_\phi)^2 A_\mu \cdot A^\mu
+\fr 1 2(\pa_\mu \eta)^2
+ \fr 1 2(\pa_\mu h)^2-\fr 1 2m^2_{\eta} \eta^2 -\fr 1 2 m^2_{h} h^2 + A_\mu \cdot A^\mu [ \kappa^\phi_\eta \eta^2  \crn
&+&\kappa_h^\phi h^2 + \kappa_{h \eta}^\phi \eta h +2 v_\phi \xi^\phi_\eta \eta + 2 v_\phi \xi^\phi_h h]
+(2 W_\mu^+ W^{-\mu} + \fr{1}{\cos^2 \theta_W} Z^\mu Z_\mu) [ \kappa_\eta \eta^2 +\kappa_h h^2  \crn
&+& \kappa_{h \eta} \eta h + 2 v \xi_\eta \eta + 2 v \xi_h h] -\la_{\eta}\eta^4 -\la_{h} h^4 -\la_1 \eta^2 h^2  -\la_2 h^3 \eta  -\la_{3} h \eta^3-\rho_{\eta} \eta^3 - \rho_{h} h^3 \crn
 &-& \rho_1 \eta^2 h - \rho_2 h^2 \eta \,,
\label{lagrfin}
\eea
where completed list of the notations is given in appendix of Ref. \cite{DarkSM-1}.

From the above Lagrangian it follows  that, beside SM parameters,  the  model has only four new parameters($m_A,m_\eta, v_\phi, g_\phi$) . Therefore it is potentially investigated at colliders.

Now we turn to the electron-positron collision in which there exists  an emission of one photon. The above process belongs to the called initial state radiation (ISR).

\section{Cross section for initial state radiation  }
\label{isrcs}
In the framework of this model,  the ISR process with  the final state consisting  of  pair of vector DM $A$ and photon is as follows
\be e^- (p_1) + e^+(p_2) \rightarrow A(p_3) + A(p_4) + \ga(p_\ga) \, .\label{n301}
\ee

\begin{figure}[]
\centering

\includegraphics[width=0.35\textwidth]{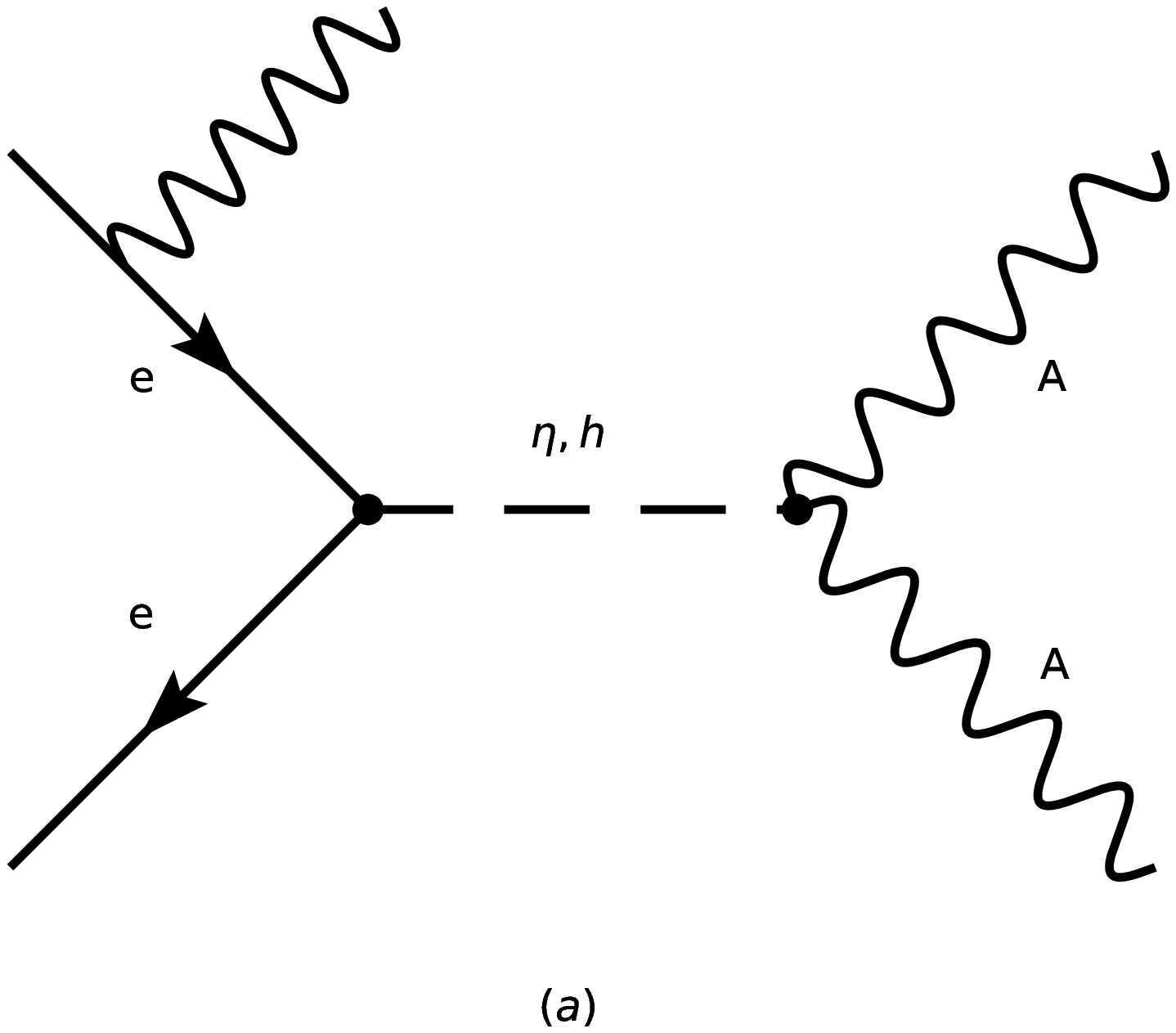}
\includegraphics[width=0.35\textwidth]{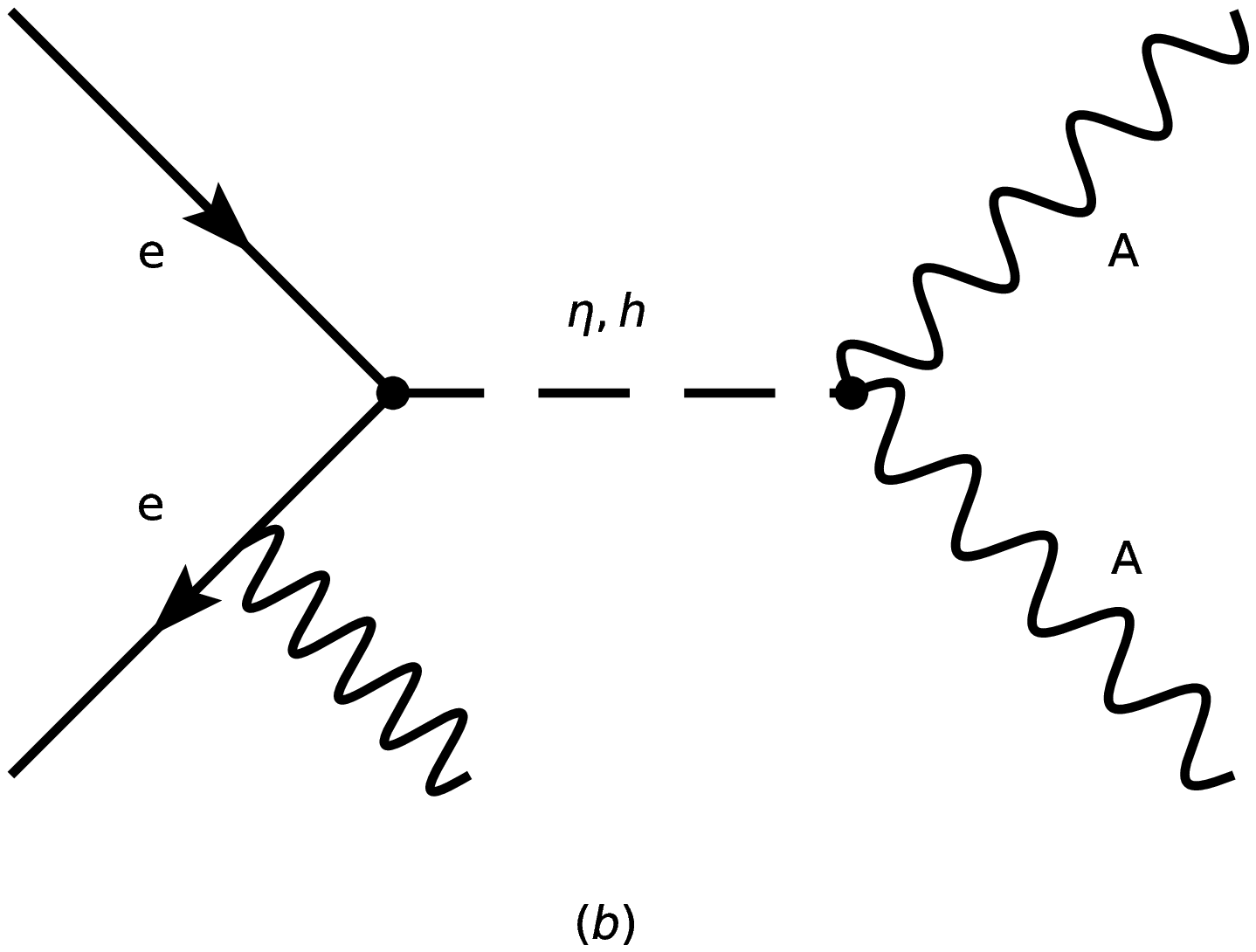}
\centering
\caption{Feynman diagram for the ISR  process $e^+e^- \rightarrow 2A+\ga$}
\label{VectorDM}       
\end{figure}

The Feynman diagrams giving main contribution to the above process
are  depicted  in Fig.\ref{VectorDM}

 Note that due to $h-\eta$ mixing, ones have coupling of the SM particles with that in the hidden sector as shown in Tables \ref{Yukawa coupling constant} and \ref{coupling constant}.

   In the energy limit where the mass of the electron can be neglected the electron propagator can be approximated as:
\be
\fr{i(p_1-  p\!\!\!/_{\ga}) +m_e }{(p_1-p_\ga)^2 -m_e^2} \approx  \fr{i(p\!\!\!/_1-  p\!\!\!/_{\ga})  }{-2p_1.p_\ga}
\label{n302}
\ee
The four momentum of the electron and photon can be written as
\be
p_1=(E,0,0,E) \hs  p_\ga=(zE,0, p_\perp, \sqrt{z^2E^2-p_\perp^2})
\label{n303}
\ee
where the photon fraction energy $z=\fr{E_\ga}{E}$ and the transverse momentum $p_\perp$. The scattering amplitude for the process in Fig. \ref{VectorDM}(a) can be written as:
\bea
 \mathcal{M}_{fi}&=&\sum_{h,\eta}\bar{v}(p_2)(-ie \ga^\mu) \mathcal{Y}^\Phi_{h, \eta}  \fr{i(p\!\!\!/_1-  p\!\!\!/_{\ga})  }{ -2p_1.p_\ga}  u(p_1) \ep _\mu^A(p_\ga) \times  \fr{i}{q^2-m^2_{h, \eta}} (ig^\Phi_{h, \eta})\ep _\al (p_3) \ep ^\al(_4).  \label{n304}
\eea
In the limit of the small   transverse momentum of the photon $ p_\perp = z p_1$, $p_\ga\approx zp_1$ the electron propagator is nearly on-shell, using the spin sum
\cite{isr}
\be
p\!\!\!/_1-p\!\!\!/_\ga \approx \sum_s u^s((1-z)p_1)\bar{u}^s((1-z)p_1)\, .
\label{n305}
\ee
Using (\ref{n305}) yields

\bea
\mathcal{M}_{e^+ e^- \rightarrow 2A + \ga } &\approx&\bar{v}(p_2)(e \ga^\mu) \mathcal{Y}^\Phi_{h, \eta}  \fr{-( \sum_s u^s((1-z)p_1)\bar{u}^s((1-z)p_1))  }{ 2p^2_\perp/2z} \crn
&& \times u(p_1) \ep _\mu^a(p_\ga)
  \cdot  \fr{i}{q^2-m^2_{h, \eta}}   (ig^\Phi_{h, \eta}) \ep _\al (p_3) \ep ^\al  (p_4)   \crn
&=&-\fr{z}{p^2_\perp}.\sum_s \mathcal{M}_{e\rightarrow e \ga}(z) \cdot
\mathcal{M}_{e^+ e^- \rightarrow 2A}((1-z)p_1, p_{2}\rightarrow p_3,p_4)
\label{n306}
\eea
where the splitting amplitude is defined as:
\be
\mathcal{M}_{e\rightarrow e \ga}(z)=e \bar{u}^s(p_1(1-z))\ga^\mu u(p_1) \ep^A_\mu(p_\ga)
\label{n307}
\ee
and the coupling constants in (\ref{n306}) are given in Tables \ref{Yukawa coupling constant} and  \ref{coupling constant}.

\begin{table}[]
\caption{Yukawa coupling constant} 
\begin{tabular}{|c| c| c| } 
\hline
Vertex  &  $\bar{f} f $ h  & $\bar{f} f  \eta $  \\
\hline 
$\mathcal{Y}^\Phi_{h, \eta}$  & $ -i\fr{m_f}{v_h} \cos{\beta} $ & $-i\fr{ m_f}{v_h} \sin{\beta} $ \\ 
\hline 
\end{tabular}
\label{Yukawa coupling constant} 
\end{table}

\begin{table}[]
\centering
\caption{ Coupling constants } 
\begin{tabular}{|c| c| c| } 
\hline
Vertex  &  $AA $ h  & $AA  \eta $  \\
\hline 
$g^\Phi_{h, \eta}$  & $ \fr{-i\eta_{\mu \nu }}{2}v_{\Phi} g^2_\phi \sin{\beta}$ &  $  \fr{i \eta_{\mu \nu }}{2} v_{\Phi} g^2_\phi \cos{\beta}$\\ 
\hline 
\end{tabular}
\label{coupling constant} 
\end{table}

The differential cross section is then be:

\bea
d\si &
\approx&\fr{1}{2s} d\Pi_\ga \left( \fr{z}{p^2_\perp} \right)^2 \fr{1}{2} \sum_{spin}|\mathcal{M}_{e\rightarrow e \ga}|^2 \fr{d^3 p_3}{(2\pi)^3}\fr{1}{2E_3}\fr{d^3 p_4}{(2\pi)^3}\fr{1}{2E_4}|\mathcal{M}_{e^+ e^- \rightarrow 2 A}|^2 \crn
&&\de ^4(p_3+p_4 -p_1 -p_2 + p_\ga)\, .
\label{dcross-section1}
\eea
The splitting amplitude can be calculated as:
\be
\fr{1}{2} \sum_{spin}|\mathcal{M}_{e\rightarrow e \ga}|^2=\fr{2e^2 p^2_\perp}{z(1-z)} \left[\fr{1+(1-z)^2}{z} \right]
\ee

The differential cross-section can be rewritten as:

\bea
\si & \approx & d\Pi_\ga \left( \fr{2e^2}{p^2_\perp} \right)^2[1+(1-z)^2] \fr{1}{2s(1-z)} \fr{d^3 p_3}{(2\pi)^3}\fr{1}{2E_3}\fr{d^3 p_4}{(2\pi)^3}\fr{1}{2E_4}|\mathcal{M}_{e^+ e^- \rightarrow 2 A}|^2\crn
&&\de ^4(p_3+p_4 -(1-z)p_1 -p_2 + p_\ga)
\label{dcross-section2}
\eea
Similarly we can write the amplitude for the process in Fig.\ref{VectorDM}(b).

The differential cross section of the process $e^+e^- \rightarrow 2A+\ga$ can be factorized as following:

\be
\fr{d\si(e^+ e^- \rightarrow 2 A + \ga)}{dz d\cos{\theta_\ga}} \approx \mathcal{F}(z, \cos{\theta_\ga}) \si(e^+ e^- \rightarrow 2 A)
\ee
where $E_\ga$ is the photon energy and $\theta_\ga$ is the angle between the photon direction and the direction of the incoming electron beam.
The collinear factor $\mathcal{F}$ is given as  \cite{isr}
\[
\mathcal{F}(x,\cos{\theta})=\fr{\al }{\pi}\fr{1+(1-x)^2}{x}\fr{1}{\sin^2{\theta}}\, .
\]
The cross-section of the process $e^+ e^- \rightarrow 2A $ is calculated as follows:
\bea
\sigma(e^+ e^- \rightarrow 2A )=\fr{ \sqrt{1-\fr{ 4 m^2_{A} }{(1-z)s} }}{16 \pi [ (1-z)s] }
 \times \fr{1}{4}  |\mathcal{M}_{fi}|^2
\eea
where
\bea
\fr{1}{4}|\mathcal{M}_{fi}|^2&=&-g_\phi^2 m_e^2  \left(-4 s m_{A}^2+12 m_{A}^4+s^2\right) \left(2 m_e^2+ s (z-1)\right) \nonumber\\
&\times &\fr{ \left( \cos{2\beta} s (z-1)+ \cos\beta^2 m_{\eta }^2-\sin\beta^2 m_h^2\right)^2+\sin\beta^4 \Ga _h^2 m_h^2}{8 v_h^2 m_{A}^2 \left(m_{\eta }^2+s (z-1)\right)^2 \left(\Ga _h^2 m_h^2+\left(m_h^2+s (z-1)\right)^2\right)}\, .
\label{Samplitude}
\eea

\section{Results}

The single photon search for new physics leads to SM background which includes mainly:

\begin{itemize}
\item Bhabha scattering of leptons with an additional  photon, $e^+ e^- \rightarrow l^+ l^- \ga$, $l= e, \mu, \tau$, where $\ga$ comes from initial state and final state radiation.
\item Neutrinos from $e^+ e^- \rightarrow \nu \bar{\nu} +\ga $
\end{itemize}

For the process $e^+ e^- \rightarrow l^+ l^- \ga$  the energy of this type of photon  however is small at order of several GeVs and the angle distribution is along the beam (forward and backward peaked). Therefore, in the event selection if the phase space cut requirement for the energy and the angle distribution is applied then the contribution of the background to the cross section is
negligible. Hence the   irreducible  background    is  the  process $e^+ e^- \rightarrow \nu \bar{\nu} +\ga $.

	In our calculation for the ISR cross-section the calculation is performed in the small energy limit of the photon and the photon is colinear. However, in an experiment, very soft and very collinear ISR photons are unobservable. Photons emerging at a small
angle to the beam line do not get detected, since  soft photons leave very small energy deposits in the calorimeter, which are drowned by noise. Thus, experimental conditions set minimal values of energy and angle for which the photon is registered.
	
	The validity of collinear approximation for cross section is discussed in \cite{DM-ILC-ModelIndependent}. They have showed that the colinear approximation for the cross section perform well for large angle and high transverse momentum photons.

	For the ILC \cite{ILC-1,ILC-2,ILC-3,ILC-4} we will use the proposed parameters of operation as follows: ILC-250 ( $\sqrt{s}$ = 250 GeV, $L_{int} = 250 fb^{-1} $);
	 ILC-500 ($\sqrt{s}$ = 500 GeV, $L_{int} = 500 fb^{-1} $); and ILC-1000 ( $\sqrt{s}$ = 1000 GeV, $L_{int} = 1000 fb^{-1} $). To select events, for all scenarios, the energy of the radiated photon $ E_\ga^{ min} = 8$ GeV  and the angle
	   $|\cos \theta_{\ga}|^{max}=0.995$.

\begin{figure}[]
\centering
\includegraphics[width=0.42\textwidth]{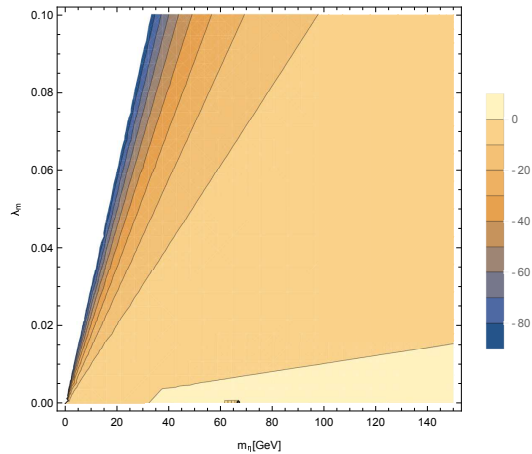}

\caption{$\tan{2\beta}$ as a function of mass of $\eta$ and mixing coupling constant.}
  \label{Tanbeta}     
\end{figure}

In our evaluation of the cross section we will first evaluate the mixing angle $\beta$. The value of mixing angle is evaluated based on the value of the mass of new scalar $\eta$ and the mixing interaction coupling constant $\la_m$ in Fig.\ref{Tanbeta}.	In most region of parameters the mixing angle is too small. In the region where the mass of $\eta$:  $ m_\eta \in [30,100] $ GeV and $\la_m \in [\times 10^{-8},   10^{-1} ]$ the mixing angle is not too small which result in large contribution to the cross section. In our  approximation where we have used the mass of the SM Higgs boson $m_h=125.10$ GEV with vacuum expectation value $v_1= 246$ GeV while the vacuum expectation value of the second Higgs $\eta$ is $v_2=5000$ GeV \cite{PDG}.

In this limit the expression (\ref{Samplitude}) becomes:
\[
 \fr{1}{4}|\mathcal{M}_{fi}|^2=\fr{-g_\phi^2 m_e^2  \left(-4 s m_{A}^2+12 m_{A}^4+s^2\right) s (z-1) }{8v_h^2 m_A^2 (m_\eta^2 +s(z-1))^2}
\]

\begin{figure}[]
\centering

\includegraphics[width=0.6\textwidth]{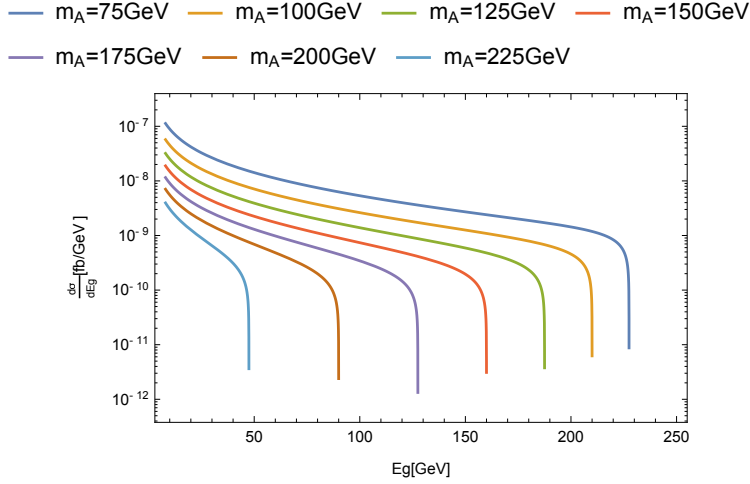}
\caption{The photon spectra from the $e^+e^- \rightarrow 2A + \ga$.}
  \label{Photon-spectra}      
\end{figure}

The shape of the photon energy spectrum for several values of the mass of the dark gauge bosons is plotted on Fig.\ref{Photon-spectra}. From the figure we can see that the photon energy spectrum almost constant for specific value of the energy of the photon.  The higher the mass $m_A$ the lower $E_{\ga,max}$. This feature the mass dependent the cut off  at the maximal allowed energy of the photon. The shape of the photon energy threshold near its endpoint where the cross section rises.

\begin{table}[]
\centering
 \begin{tabular}{|c|| c| c| c| c| c| c| c|}
 \hline
 $M$(GeV) & 75 & 100 & 125 & 150 & 175 & 200 & 225 \\ [0.5ex]
 \hline\hline
 $\sigma_{bg}(fb)$ & 36 & 83 & 202 & 590 & 2030 & 1800 & 1200 \\
 \hline
 $\sigma_{WIMP}(fb)$ & 1.8 & 3.9 & 8.4  & 21 & 64 & 27 & 4.9\\
 \hline
 $\sigma_{A}$ $\times 10^{6}(fb)$ & 2.24 & 1.1 & 0.54 & 0.31 & 0.186 & 0.089 & 0.04\\
 \hline
 \end{tabular}
 \caption{Signal and background cross sections at $\sqrt{s}$ = 500 GeV with no polarization and the cuts specified in the text, for
a few representative values of dark gauge bosons $M_A$. }
\label{Cross-section-table}
\end{table}

The cross section for different value of the mass are given in Table \ref{Cross-section-table}. We evaluated the cross-section with different values of mass of hidden-sector gauge bosons at center of mass $\sqrt{s}=500$ GeV to compare with the background and the WIMPs signal  calculated in \cite{DM-ILC-ModelIndependent}.

 These values are of order $10^{-44} cm^2$ which equals to that of direct detection \cite{DarkSM-1} and can be comparable with current experiment \cite{CDMS}. However these values are  significantly smaller than the background therefore further analysis and methods to suppress the background is needed.

\begin{figure}[]
\centering
\includegraphics[width=0.5\textwidth]{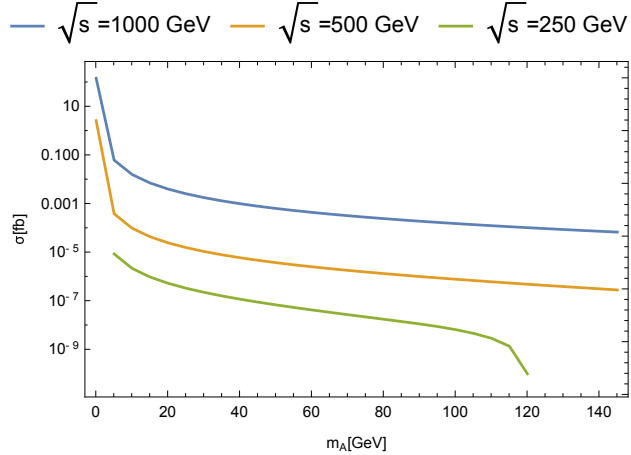}
\caption{Cross section as a function of mass of dark gauge bosons for different value of center of mass energy for process $e^+e^- \rightarrow 2A + \ga$.}
  \label{Cross-section-mA}      
\end{figure}

In Fig. \ref{Cross-section-mA} we investigate the cross section with the mass of the dark gauge bosons for different values of the center of mass energy. The energy of the emitted photon is integrated from 8 GeV to 50 GeV.

\begin{figure}[]
\centering
\includegraphics[width=0.5\textwidth]{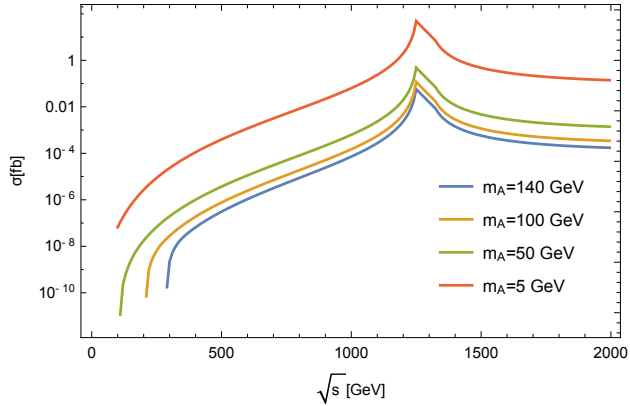}
\caption{Cross section vs $\sqrt{s}$ with different value of mass of dark gauge bosons for process $e^+e^- \rightarrow 2A + \ga$.}
  \label{Cross-section-S}     
\end{figure}

In Fig. \ref{Cross-section-S} we investigate the cross section with $\sqrt{s}$ for several values of the mass of the dark gauge bosons. If the mass of the dark gauge boson is about 5 GeV then there exits a resonance where the signal cross section can be as large as order $48 fb$ which is  greater the back ground  \cite{DM-Collider-EffOperator} at $\sqrt{s}\approx 500$ GeV and mass of WIMPs equals 75 GeV.

\section{Discussion and conclusions}

Dark matter question is one of the great puzzle in particle physics. The origins and properties of DM are still unclear. For long time WIMPs are the most promising candidate for DM however this assumption is in checked \cite{hidden-Sector1}. Hence
DM might belong to a hidden sector with
new particles without SM quantum number.

Previous studies  at colliders often use effective approach for discussing the prospects for discovery of WIMPs dark matter \cite{DM-ILC-ModelIndependent,Iluminating-DM-ILC,DM-Collider-EffOperator}. In this work we have investigated the prospect of detecting the dark   gauge boson of the hidden $SU(2)$ extension of the SM  at ILC. The mixing of the SM  Higgs with the scalar of the new dark gauge group leads to a portal in which the dark gauge bosons interact with matter. However  due to small Yukawa coupling the cross section of the process $e^+ e^- \rightarrow \ga + 2 A$ is significantly smaller than the SM background ($e^+ e^- \rightarrow \ga \bar{\nu} \nu $). When increase the center of mass energy, the cross section is significantly  higher and can be of an order of the irreducible background and as large as the cross section of searches for WIMPs with mass 75 GeV.  Therefore, to have deeper insight into this search more work needed to be done. Specifically,  evaluate the irreducible of the background and WIMPs signal at the mass of WIMPs equal to 5 GeV  in order to compare with cross section of the hidden gauge bosons. Furthermore, to  detect hidden $SU(2)_L$ gauge bosons, more methods needed to suppress the background and increase the energy of the ILC.

\section*{Acknowledgment}
This research is funded by Vietnam  National Foundation for Science and Technology Development (NAFOSTED) under grant number 103.01-2017.356.
\\[0.3cm]

\end{document}